# Analysis of Different Approaches of Parallel Block Processing for K-Means Clustering Algorithm

Rashmi C[a]

[a]High-Performance Computing Project, Department of Studies in Computer Science, University of Mysore, Mysuru 570 006, India, Contact: rashmi.hpc@gmail.com

Abstract: Distributed Computation has been a recent trend in engineering research. Parallel Computation is widely used in different areas of Data Mining, Image Processing, Simulating Models, Aerodynamics and so forth. One of the major usage of Parallel Processing is widely implemented for clustering the satellite images of size more than dimension of 1000x1000 in a legacy system. This paper mainly focuses on the different approaches of parallel block processing such as row-shaped, column-shaped and square-shaped. These approaches are applied for classification problem. These approaches is applied to the K-Means clustering algorithm as this is widely used for the detection of features for high resolution orthoimagery satellite images. The different approaches are analyzed, which lead to reduction in execution time and resulted the influence of improvement in performance measurement compared to sequential K-Means Clustering algorithm.

Keywords: Approaches, K-Means, Parallel Processing, Satellite Images.

## 1. INTRODUCTION

Image Classification plays a major role in processing the remotely sensed images. The intent of classifying the images is to group the pixels based on similar properties [1]. Satellite image data sets are of size in Kilobytes, Megabytes and Gigabytes are common in processing the image operations for classification. K-means is an unsupervised classification method is most popularly used for image analysis, Bioinformatics, Pattern Recognition and Statistical Data analysis [2]. One of major usage of clustering is the classification of satellite images. Since the time complexity of sequential K means clustering is higher for processing the high resolution satellite images whose dimension greater than 1000x1000, hence it is considered for parallel processing.

Parallel Block Processing exhibits single program multiple data (SPMD) [3] parallel programming model, which allows for a greater control over the parallelization of tasks. The tasks could be distributed and assigned to processes or labs or sessions. An operation in which an image is processed in blocks at once. Some operation is applied to blocks parallely at a time by distributing an operation/task among the workers hence the name task parallelism. Then blocks are reassembled to form an output image. Existing clustering algorithms exhibits high computational time for larger images of pixel dimension greater than 1000. The main focus of this paper is to tackle this problem by using block processing which performs in parallel using Matlab Programming environment. The main advantage is to make use of current multi-core architectures available in commercial processors efficiently which in turn increases speedup of clustering process for the larger images. Therefore, the presented approach doesn't require special hardware and can run on machines that are commercially available.

An Efficient parallel Block processing algorithm is designed in this paper for the reduction in the processing time of satellite images of dimension more than 1000x1000 which lead to maximum usage of CPU in standalone systems. In this paper three approaches of block considered such as row-shaped, column-shaped and square-shaped. These approaches are experimentally analyzed and studied. The experimental results of different approaches of block processing is illustrated in the following sections. Hence the proposed algorithm is more efficient for processing the satellite images.

The remainder of the paper is organized as section 2 depicts the literature survey, section 3 explains about the different approaches of parallel block processing such as row-shaped, column-shaped and square-shaped followed by the experimental results comparing the serial execution time with different approaches of parallel block processing with its performance.

## 2. LITERATURE SURVEY

Jerril Mathson Mathew, Jyothis Joseph [4] exploited the usage of Hadoop for the implementation of K-Means algorithm. Map Reduce Programming Model is used for clustering algorithm. Clustering is one of the momentous task in data mining. In this paper authors describe the basic components of Hadoop platform [5] workflow of all stages of Map-Reduce including the structural relationship of HDFS framework. Manasi N. Joshi [6] exploited data parallelism for the clustering K-Means clustering algorithm with the usage of message passing model. This paper mainly focuses on the splitting and distributing data sets among processes for computation. Shared memory programming model a message passing interface (MPI) [7] has been employed for computation of vector cluster membership for each partition generated and is experimented in Sun Workstation cluster. S. Mohanvalli, S.M. Jaisakthi and C. Aravindan explored partitional K-means clustering algorithm for the current parallel architecture [8]. Their work focuses on Hybrid model of both OpenMP [9] shared memory programming model and MPI distributed memory model programming. OpenMP an application programming interface is utilized for performing task parallel and even distribution of data points among the processors, updation of new centroids and labelling of data by message passing among the processes. It is experimental shown that hybrid model for scalability and feasibility of parallelism of the algorithm was efficient than other approaches. Fahim Ahmed M proposed about the design and implementation of parallel K-Means algorithm paralyzed algorithm by using parallel for the distance computation concurrently on multicore architecture. [10] Employed parallel for loops (parfor) in matlab, where n number of iterations run on cluster of matlab workers, n/m iterations of the loop are executed by each worker. [11] Proposed algorithm is implemented for biomedical research area where the image is clustered into 2 using modified k-means. Each clustered images are sent to slave processors, processed sub-images are sent back to the master processor and results are consolidated at the master processor.

3. APPROACHES OF PARALLEL BLOCK PROCESSING

The processing time and computation power is very high for a large amount of data to be processed on all levels of image processing. Parallel Processing [12] significantly reduces the processing time for the image data, particularly multicores can be used efficiently. Some image processing operations involve processing an image in sections called blocks, rather than processing an entire image at once. The input image is divided into blocks, each block calls the specified function and results are consolidated to form an output image. Two operations to perform for block processing such as sliding neighborhood and distinct block operations. In distinct block operation, the input image is divided into block as two, four, and eight as tiles of specified dimension based on the size of an input image. Different thread processes each block parallel at a time performing K-means clustering.

In general, using larger blocks while block processing an image results in faster performance than completing the same task using smaller blocks. However, sometimes the task or algorithm applied to an image requires a certain block size, and smaller blocks need to be used. When block processing using smaller blocks, parallel block processing is typically faster than regular (serial) block processing, often by a large margin. Image Size, Block Size and function used for processing are the factors need to be considered for the parallel block processing. Time consumption mainly depends on the selection of block size resulting in improvement of the performance measurement. The three approaches [13] result the influence of performance measurement are Row-Shaped, Column-Shaped and Square Block as illustrated in Figure 2. Figure 1 illustrates the proposed block diagram of Parallel Processing for clustering the orthoimagery satellite images.

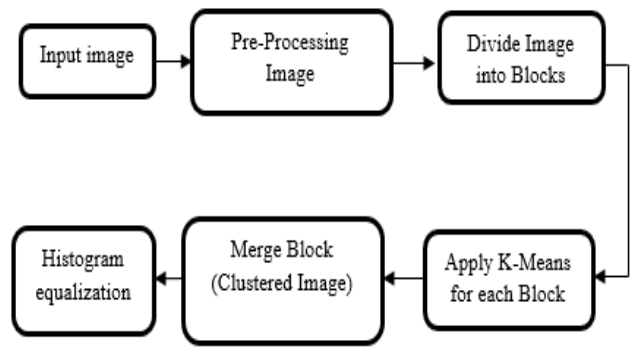

Fig .1 Block Diagram of proposed system

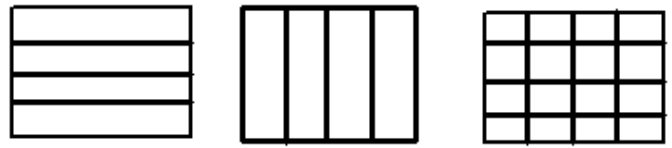

Fig. 2 Approaches of Block Processing Row-Shaped, Column-Shaped and Square Block

## 4. EXPERIMENTAL RESULTS

Results obtained with the parallel block processing of high resolution orthoimages are presented followed by the performance evaluation.

The testing is done on commercially available systems of hardware Intel Xeon CPU E3-1220 V2@3.10GHz 8.00 GB RAM. The proposed system is implemented in Matlab 2014a programming environment by varying block size and number of workers as 2, 4 and 8 with the consideration of clusters as 2 and 4. Two datasets comprising of more than 100 samples of medium and high resolution [14] aerial images is considered for K-means clustering. For Medium resolution images of 8-bit depth of resolution 30,50 and 40cm where file size of 786KB,955KB,1.17MB of pixel dimension 2000x1024 is considered. For High resolution images 16-bit depth of resolution 30, 50 and 80 cm, file size exceeding 800 KB but limited to 1GB, image pixel dimensions of 1280x800, 2640x2640, 4656x5793, 5528x5350 and 9052 x 4965 covering large area of approximately 2.8x3.0 Km of some parts of India, USA, chile, France and so forth with 3 spectral bands (RGB) is considered for the experimentation. Figure 3 illustrates sample input image of high resolution orthoimagery. Figure 4 depicts the output of Sequential K-Means for Cluster 2 and Figure 5 illustrates the Parallel Block Processing for cluster 2. Figure 6 depicts Sequential K-Means for Cluster 4 and Figure 7 illustrates the Parallel Block Processing for Cluster 4.

The performance is affected by the number of times the file is accessed to either read or write image files when using blockproc function. Selecting the larger block size reduces the number of times blockproc has to access the disk, at the cost of using more memory to process each block. Influence of block size on the performance of blockproc is demonstrated by the following three cases. In each of the cases, the number of pixels in each block is approximately the same only the size of blocks is different. An image of pixel dimension 4656x5793 of 77.3MB is considered for the illustration of performance of block size for cluster 2 with its execution time.

Case 1: Typical case — Square-Block
A square block of size [1200 1200] is considered. The image is 4656 pixels wide and 1200 rows wide and it's approximately (4656/1200=3.88). Since the image is 4 blocks wide, blockproc reads every strip of the file 4 times and its elapsed time is 0.256ms (milliseconds) for worker 2, 0.147ms for worker 4 and 0.143ms for worker 8 respectively.

Case 2: Worst case — Row-Shaped Block
A Block Size [1200 4656] is chosen. Each spans the width of an image. Each strip is read spans the width of an image. Each strip is read exactly one time and all data for a particular block is stored continuously on disk and its execution time is 0.249ms for worker 2, 0.146ms for worker 4 and 0.144ms for worker 8.

Case 3: Best case — Column-Shaped Block
Blocks shaped like columns of size [5793 1000] is chosen. The image is over 4.656 approximately 5 blocks wide (4656/1000=4.656). The block proc function reads an entire image from disk 5 times. The elapsed is 0.244ms for worker 2, 0.140ms for worker 4 and 0.144ms for worker 8.

From the Experimental results, it is evident that the Column-Shaped is considered to be the best case comparable to other approaches of parallel block processing.

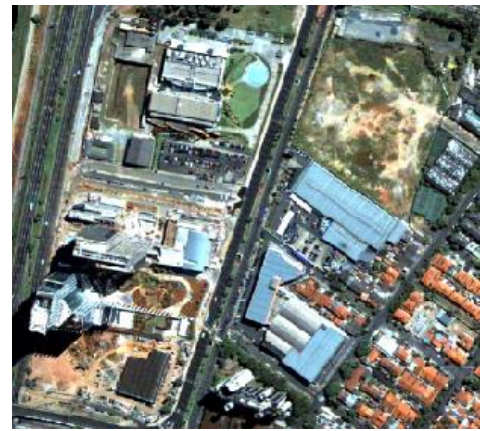

Fig 3. Input Image

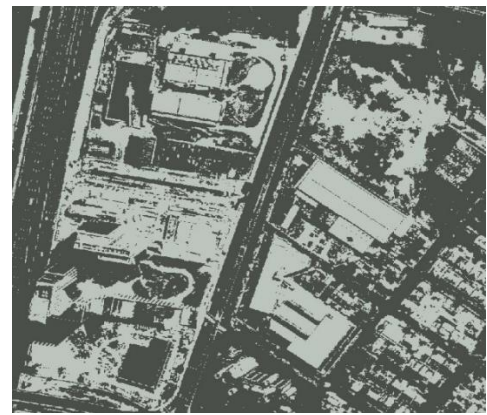

Fig 4. Sequential K-Means for Cluster 2

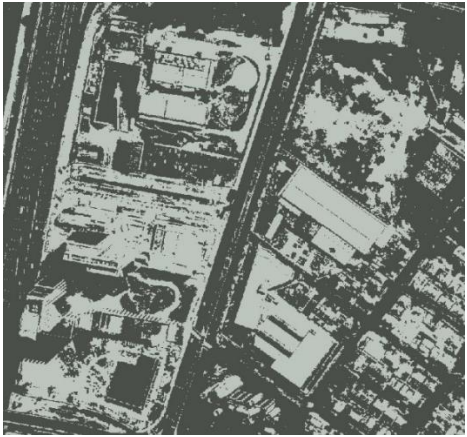

Fig 5. Parallel Block Processing for cluster 2

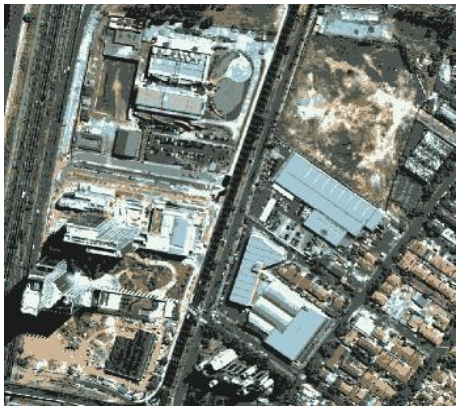

Fig 6. Sequential k-Means for Cluster 4

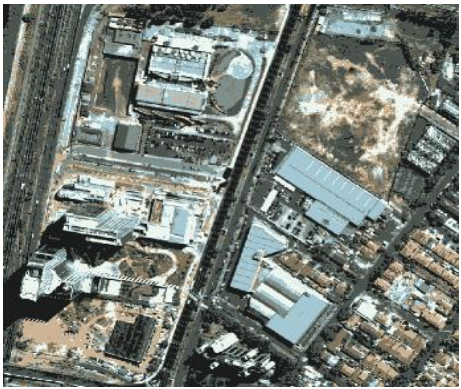

Fig 7. Parallel Block Processing for cluster 4

4.1 Performance Evaluation

An evaluation of parallel block processing is done for its performance study by varying the number of cores such as 2, 4 and 8 for the cluster 2 and 4. Different approaches of Parallel Block Processing such as Row-Shaped, Column-Shaped and Square Block are illustrated in table followed by its graphical representation.

Table 1 and Table 2 illustrates the Speedup calculation for Row-Shaped, Cluster 2 for 2 Cores and for 4 Cores. followed by it's the graphical representation in Figure 8 and figure 9.

Table 1. Efficiency calculation for Row-Shaped, Cluster 2, 2 cores

| Efficiency Calculation for Cluster 2 2 Cores, Time (ms) | | | | |
|---|---|---|---|---|
| Data Size | Serial | Parallel | Speedup | Efficiency |
| 1024x768 | 0.050589 | 0.036366 | 1.391107078 | 0.695553539 |
| 1226x878 | 0.056069 | 0.048666 | 1.152118522 | 0.576059261 |
| 3729x2875 | 0.591048 | 0.39576 | 1.493450576 | 0.746725288 |
| 1355x1255 | 0.091383 | 0.064567 | 1.41532052 | 0.70766026 |
| 5528x5350 | 1.895121 | 1.054863 | 1.79655652 | 0.89827826 |
| 2640x2640 | 0.437126 | 0.24845 | 1.759412357 | 0.879706178 |
| 4656x5793 | 1.714137 | 0.249265 | 6.876765691 | 3.438382846 |
| 5490x5442 | 1.971303 | 0.264342 | 7.457396101 | 3.72869805 |
| 9052x4965 | 2.442462 | 1.994543 | 1.224572245 | 0.612286123 |

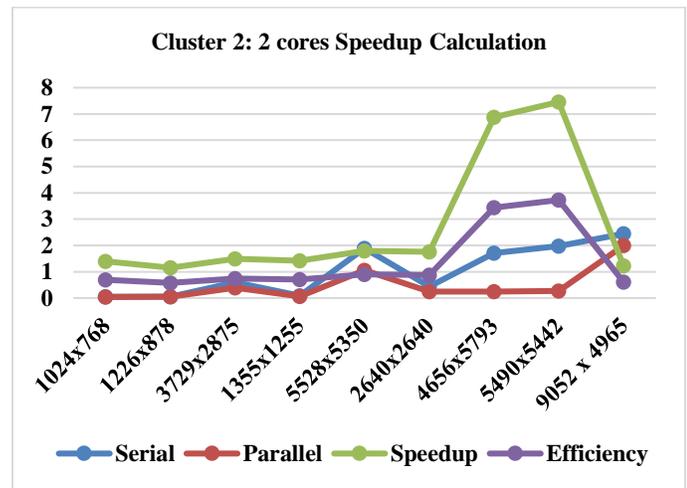

Fig 8. Speed Up for row-shaped, Cluster 2, 2 Cores

Table 2. Efficiency calculation for Row-Shaped cluster 2, 4 cores

| Efficiency Calculation for Cluster 2 4 Cores, Time(ms) | | | | |
|---|---|---|---|---|
| Data Size | Serial | Parallel | Speedup | Efficiency |
| 1024x768 | 0.050589 | 0.016993 | 2.977049373 | 0.744262343 |
| 1226x878 | 0.056069 | 0.02372 | 2.363785835 | 0.590946459 |
| 3729x2875 | 0.591048 | 0.38743 | 1.525560746 | 0.381390187 |
| 1355x1255 | 0.091383 | 0.036209 | 2.52376481 | 0.630941202 |
| 5528x5350 | 1.895121 | 0.622445 | 3.044640089 | 0.761160022 |
| 2640x2640 | 0.437126 | 0.153703 | 2.84396531 | 0.710991327 |
| 4656x5793 | 1.714137 | 0.144857 | 11.83330457 | 2.958326142 |
| 5490x5442 | 1.971303 | 0.152811 | 12.90026896 | 3.22506724 |
| 9052x4965 | 2.442462 | 1.286078 | 1.899155417 | 0.474788854 |
| 2640x2640 | 0.437126 | 0.24521 | 1.782659761 | 0.122605 |
| 4656x5793 | 1.714137 | 0.244717 | 7.004568542 | 0.1223585 |
| 5490x5442 | 1.971303 | 0.255725 | 7.708683156 | 0.1278625 |
| 9052x4965 | 2.442462 | 1.817349 | 1.343969705 | 0.9086745 |

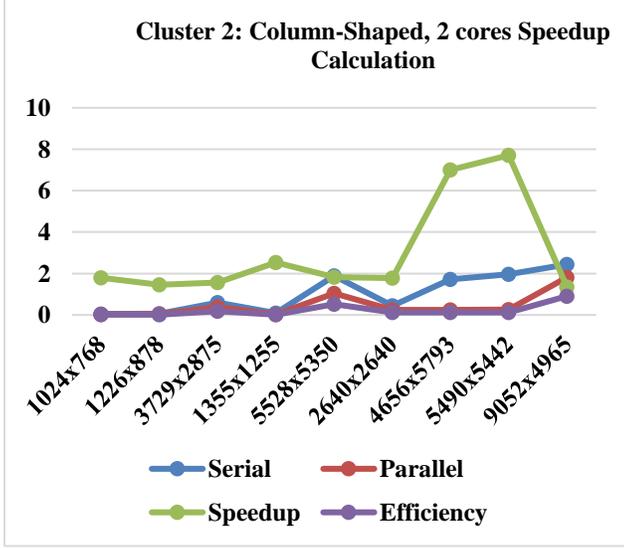

Fig 10. Speed Up for Column-Shaped, cluster 2, 2 cores

Table 4. Efficiency calculation for Column-Shaped, Cluster 2, 4 Cores

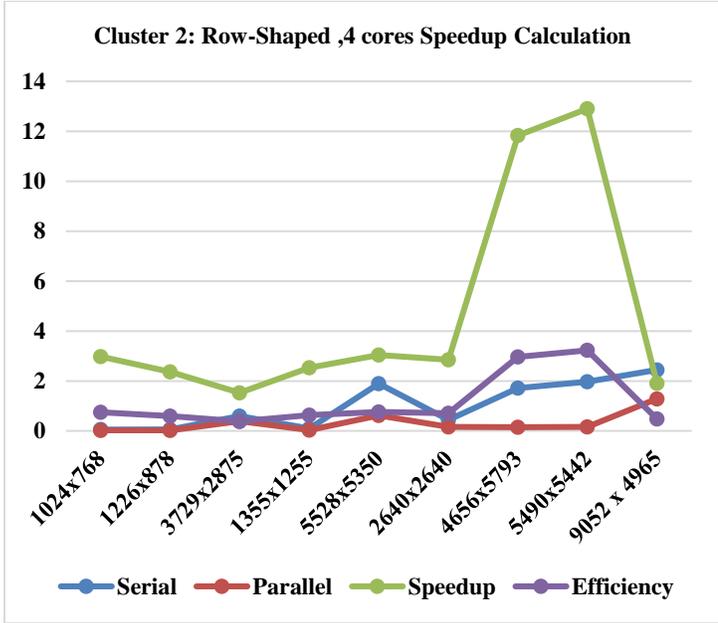

Fig 9. Speed Up for Row-Shaped, Cluster 2, 4 Cores

Table 3 and Table 4 illustrates the Speedup calculation for Column-Shaped, Cluster 2 for 2 Cores and for 4 Cores. followed by it's the graphical representation in Figure 10 and figure 11.

Table 3. Efficiency calculation for Column Shaped, Cluster 2, 2 Cores

| Efficiency Calculation for Cluster 2 | | | | |
|---|---|---|---|---|
| 2 Cores, Time(ms) | | | | |
| Data Size | Serial | Parallel | Speedup | Efficiency |
| 1024x768 | 0.050589 | 0.02808 | 1.801602564 | 0.01404 |
| 1226x878 | 0.056069 | 0.03857 | 1.453694581 | 0.019285 |
| 3729x2875 | 0.591048 | 0.378012 | 1.56356941 | 0.189006 |
| 1355x1255 | 0.091383 | 0.035979 | 2.539898274 | 0.0179895 |
| 5528x5350 | 1.895121 | 1.041893 | 1.818920945 | 0.5209465 |

| Efficiency Calculation for Cluster 2 | | | | |
|---|---|---|---|---|
| 4 Cores, Time(ms) | | | | |
| Data Size | Serial | Parallel | Speedup | Efficiency |
| 1024x768 | 0.050589 | 0.0161 | 3.142173913 | 0.785543478 |
| 1226x878 | 0.056069 | 0.02185 | 2.566086957 | 0.641521739 |
| 3729x2875 | 0.591048 | 0.22463 | 2.631206874 | 0.657801718 |
| 1355x1255 | 0.091383 | 0.33547 | 0.272402897 | 0.068100724 |
| 5528x5350 | 1.895121 | 0.613971 | 3.086662074 | 0.771665518 |
| 2640x2640 | 0.437126 | 0.141992 | 3.078525551 | 0.769631388 |
| 4656x5793 | 1.714137 | 0.140939 | 12.16226169 | 3.040565422 |
| 5490x5442 | 1.971303 | 0.143862 | 13.70273596 | 3.425683989 |
| 9052x4965 | 2.442462 | 0.940208 | 2.597789 | 0.64944725 |

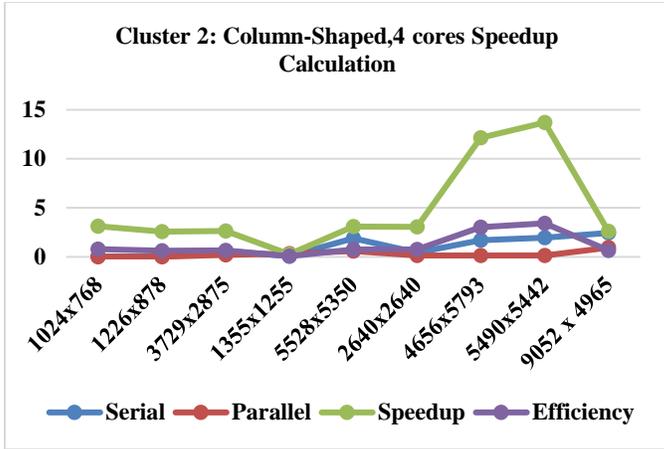

Fig 11. Speed Up for Column-Shaped, cluster 2, 4 cores

Table 5 and Table 6 illustrates the Speedup calculation for Square Block, Cluster 2 for 2 Cores and for 4 Cores. followed by it's the graphical representations.

Table 5. Efficiency Calculation for Square Block, Cluster 2, 2 cores

| Efficiency Calculation for Cluster 2 2 Cores, Time(ms) | | | | |
|---|---|---|---|---|
| Data Size | Serial | Parallel | Speedup | Efficiency |
| 1024x768 | 0.050589 | 0.03236 | 1.563318912 | 0.781659456 |
| 1226x878 | 0.056069 | 0.039234 | 1.429092114 | 0.714546057 |
| 3729x2875 | 0.591048 | 0.39141 | 1.510048287 | 0.755024143 |
| 1355x1255 | 0.091383 | 0.061011 | 1.49781187 | 0.748905935 |
| 5528x5350 | 1.895121 | 0.588896 | 3.21809114 | 1.60904557 |
| 2640x2640 | 0.437126 | 0.257618 | 1.696799137 | 0.848399568 |
| 4656x5793 | 1.714137 | 0.256567 | 6.681050174 | 3.340525087 |
| 5490x5442 | 1.971303 | 0.257681 | 7.650168231 | 3.825084116 |
| 9052x4965 | 2.442462 | 2.286979 | 1.067986195 | 0.533993097 |

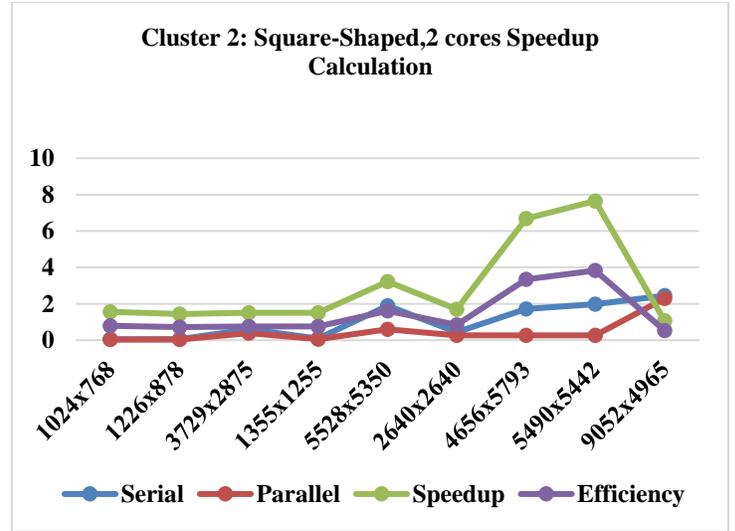

Fig 12. Speed Up for Square Block, cluster 2, 2 cores

Table 6. Efficiency Calculation for Square Block, Cluster 2, 4 Cores

| Efficiency Calculation for Square Block, Cluster 2 4 Cores, Time(ms) | | | | |
|---|---|---|---|---|
| Data Size | Serial | Parallel | Speedup | Efficiency |
| 1024x768 | 0.050589 | 0.02029 | 2.493297191 | 0.623324298 |
| 1226x878 | 0.056069 | 0.021877 | 2.562919962 | 0.64072999 |
| 3729x2875 | 0.591048 | 0.22142 | 2.669352362 | 0.667338091 |
| 1355x1255 | 0.091383 | 0.034796 | 2.626250144 | 0.656562536 |
| 5528x5350 | 1.895121 | 0.622445 | 3.044640089 | 0.761160022 |
| 2640x2640 | 0.437126 | 0.150379 | 2.906828746 | 0.726707187 |
| 4656x5793 | 1.714137 | 0.14723 | 11.64257964 | 2.910644909 |
| 5490x5442 | 1.971303 | 0.147826 | 13.33529284 | 3.333823211 |
| 9052 x 4965 | 2.442462 | 2.031543 | 1.202269408 | 0.300567352 |

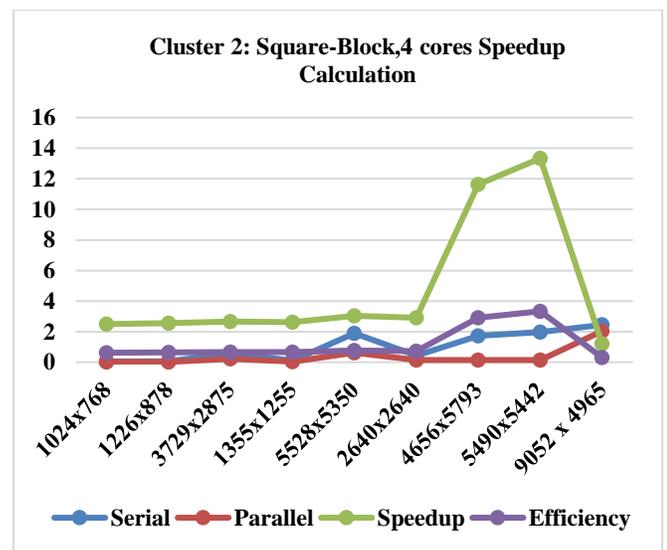

Fig 13. Speed Up calculation for Square Block, 4 Cores

Table 7 and Table 8 illustrates the Speedup calculation for Row-Shaped, Cluster 4 for 2 Cores and for 4 Cores. followed by it's the graphical representation.

Table 7. Efficiency Calculation for Row-Shaped, Cluster 4, 2 Cores

| Efficiency Calculation for Cluster 4 | | | | |
| --- | --- | --- | --- | --- |
| 2 Cores, Time (ms) | | | | |
| Data Size | Serial | Parallel | Speedup | Efficiency |
| 1024x768 | 0.0505 | 0.0458 | 1.102620087 | 0.551310044 |
| 1226x878 | 0.056 | 0.0488 | 1.147540984 | 0.573770492 |
| 3729x2875 | 0.691 | 0.5244 | 1.317696415 | 0.658848207 |
| 1355x1255 | 0.0913 | 0.0775 | 1.178064516 | 0.589032258 |
| 5528x5350 | 1.895 | 1.3811 | 1.372094707 | 0.686047354 |
| 2640x2640 | 0.537 | 0.3375 | 1.591111111 | 0.795555556 |
| 4656x5793 | 2.767 | 1.5971 | 1.732515184 | 0.866257592 |
| 5490x5442 | 3.161 | 1.7588 | 1.797248124 | 0.898624062 |
| 9052x4965 | 4.69 | 2.5893 | 1.811300351 | 0.905650176 |

Table 8. Efficiency Calculation for Row-Shaped, Cluster 4, 4 Cores

| Efficiency Calculation for Cluster 4 | | | | |
| --- | --- | --- | --- | --- |
| 4 Cores, Time(ms) | | | | |
| Data Size | Serial | Parallel | Speedup | Efficiency |
| 1024x768 | 0.0505 | 0.029596 | 1.706311664 | 0.426577916 |
| 1226x878 | 0.056 | 0.039824 | 1.406187224 | 0.351546806 |
| 3729x2875 | 0.691 | 0.469335 | 1.472295908 | 0.368073977 |
| 1355x1255 | 0.0913 | 0.04888 | 1.867839607 | 0.466959902 |
| 5528x5350 | 1.895 | 0.840046 | 2.255828847 | 0.563957212 |
| 2640x2640 | 0.537 | 0.232365 | 2.311019302 | 0.577754825 |
| 4656x5793 | 2.767 | 0.793099 | 3.488845655 | 0.872211414 |
| 5490x5442 | 3.161 | 0.892304 | 3.542514659 | 0.885628665 |
| 9052x4965 | 4.69 | 1.340988 | 3.497421304 | 0.874355326 |

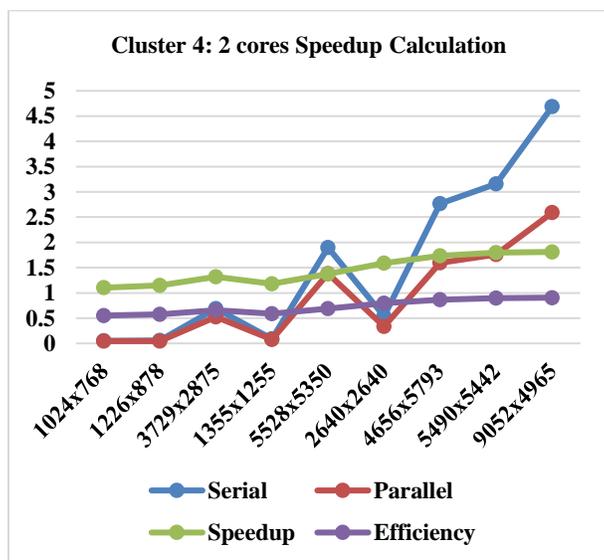

Fig 14. Speed Up Calculation of Row-Shaped, cluster 4, 2 cores

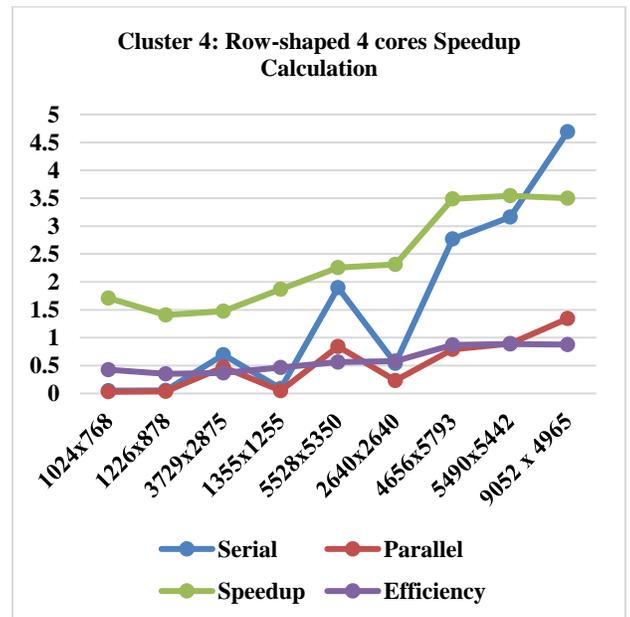

Fig 15. Speed Up Calculation for Row-Shaped, Cluster 4, 4 Cores

Table 9. Efficiency Calculation for Column-Shaped, Cluster 4, 4 Cores

| Efficiency Calculation for Column-Shaped, Cluster 4 4 Cores, Time(ms) | | | | |
|---|---|---|---|---|
| DataSize | Serial | Parallel | Speedup | Efficiency |
| 1024x768 | 0.050589 | 0.02211 | 2.288059701 | 0.572014925 |
| 1226x878 | 0.056069 | 0.16777 | 0.334201586 | 0.083550396 |
| 3729x2875 | 0.691048 | 0.268484 | 2.573888947 | 0.643472237 |
| 1355x1255 | 0.091383 | 0.04289 | 2.130636512 | 0.532659128 |
| 5528x5350 | 1.895121 | 0.804428 | 2.355861556 | 0.588965389 |
| 2640x2640 | 0.537126 | 0.188548 | 2.84874939 | 0.712187348 |
| 4656x5793 | 2.767155 | 0.782554 | 3.536056298 | 0.884014074 |
| 5490x5442 | 3.161864 | 0.849109 | 3.723743359 | 0.93093584 |
| 9052x4965 | 4.69014 | 1.326101 | 3.53678943 | 0.884197358 |

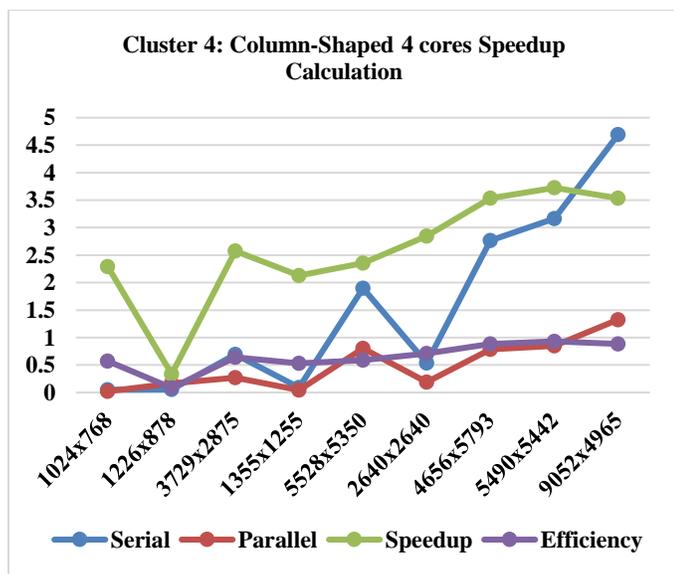

Fig 16. Speedup Calculation for Column shaped, Cluster 4, 4 Cores

Table 10. Efficiency Calculation for Square Block, Cluster 4, 4 Cores

| Efficiency Calculation for Square Block Cluster 4 4 Cores, Time(ms) | | | | |
|---|---|---|---|---|
| Data Size | Serial | Parallel | Speedup | Efficiency |
| 1024x768 | 0.050589 | 0.22416 | 0.225682548 | 0.056420637 |
| 1226x878 | 0.056069 | 0.027785 | 2.017959331 | 0.504489833 |
| 3729x2875 | 0.691048 | 0.372608 | 1.854624699 | 0.463656175 |
| 1355x1255 | 0.091383 | 0.048753 | 1.874407729 | 0.468601932 |
| 5528x5350 | 1.895121 | 0.805928 | 2.351476807 | 0.587869202 |
| 2640x2640 | 0.537126 | 0.235777 | 2.278110248 | 0.569527562 |
| 4656x5793 | 2.767155 | 0.800618 | 3.456273779 | 0.864068445 |
| 5490x5442 | 3.161864 | 0.868938 | 3.638768244 | 0.909692061 |
| 9052x4965 | 4.69014 | 1.328736 | 3.529775666 | 0.882443917 |

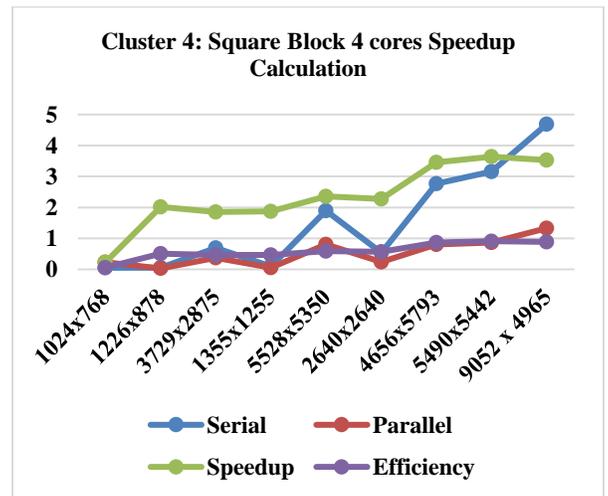

Figure 17. Speed Up Calculation for Square Block, Cluster 4, 4 Cores

Table 11. Efficiency Calculation for Square Block, Cluster 4, 8 Cores

| Efficiency Calculation for Square Block Cluster 4 8 Cores, Time(ms) | | | | |
|---|---|---|---|---|
| Data Size | Serial | Parallel | Speedup | Efficiency |
| 1024x768 | 0.050589 | 0.011692 | 4.326804653 | 0.540850582 |
| 1226x878 | 0.056069 | 0.014881 | 3.767824743 | 0.470978093 |
| 3729x2875 | 0.691048 | 0.218793 | 3.158455709 | 0.394806964 |
| 1355x1255 | 0.091383 | 0.027417 | 3.333078017 | 0.416634752 |
| 5528x5350 | 1.895121 | 0.413192 | 4.586538462 | 0.573317308 |
| 2640x2640 | 0.537126 | 0.129354 | 4.152372559 | 0.51904657 |
| 4656x5793 | 2.767155 | 0.427864 | 6.467370473 | 0.808421309 |
| 5490x5442 | 3.161864 | 0.45649 | 6.926469364 | 0.865808671 |
| 9052x4965 | 4.69014 | 0.68592 | 6.837736179 | 0.854717022 |

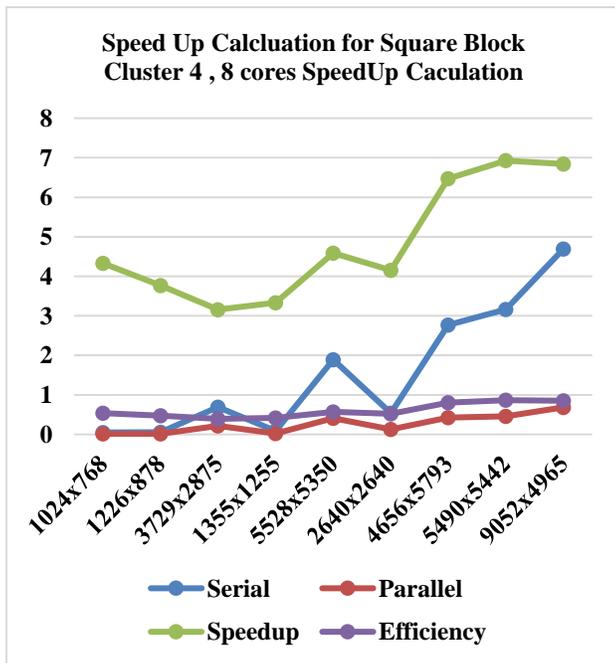

Fig. 18 Speed Up Calculation for Square Block, Cluster 4,8 Cores

Table 12 to Table 14 illustrates the results of Sequential with parallel considering the number of cores/workers as 2, 4 and 8 for the image of resolution 4656 x 5793 for Row-Shaped, Column-Shaped and Square Block for Cluster 2.

Table 12. Comparison results of Row-Shaped

| Data Size | Serial | Cores | Row-Shaped | Speed Up | Efficiency |
|---|---|---|---|---|---|
| 4656x5793 | 1.714137 | 2 | 0.249265 | 6.876 | 3.438382846 |
| 4656x5794 | 1.714137 | 4 | 0.144857 | 11.833 | 2.958326142 |
| 4656x5795 | 1.714137 | 8 | 0.146973 | 11.662 | 1.457867261 |

Table 13. Comparison results of Column-Shaped

| Data Size | Serial | Cores | Column-Shaped | Speed Up | Efficiency |
|---|---|---|---|---|---|
| 4656x5793 | 1.714137 | 2 | 0.244717 | 7.004568542 | 3.502284271 |
| 4656x5794 | 1.714137 | 4 | 0.140939 | 12.16226169 | 3.040565422 |
| 4656x5795 | 1.714137 | 8 | 0.144902 | 11.82962968 | 1.47870371 |

Table 14. Comparison results of Square Block

| Data Size | Serial | Cores | Square Block | Speed Up | Efficiency |
|---|---|---|---|---|---|
| 4656x5793 | 1.714137 | 2 | 0.256567 | 6.681050174 | 3.340525087 |
| 4656x5794 | 1.714137 | 4 | 0.14723 | 11.64257964 | 2.910644909 |
| 4656x5795 | 1.714137 | 8 | 0.143322 | 11.96004103 | 1.495005128 |

Table 15 illustrates the comparison results of different approaches of Block processing for cluster 2 followed by its graphical representation.

Table 15. Comparison of Different Approaches of Block processing for cluster 2

| Block Size | Non Block | [1200 4656] Row-Shaped | [5793 1000] Column-Shaped | [1200 1200] Square Block |
|---|---|---|---|---|
| Processing Time | 1.7141 | 0.541095 | 0.530558 | 0.547119 |
| Speed Up | | 10.124336 | 10.332153 | 10.0945569 |
| Efficiency | | 2.6181920 | 2.6738511 | 2.58205837 |

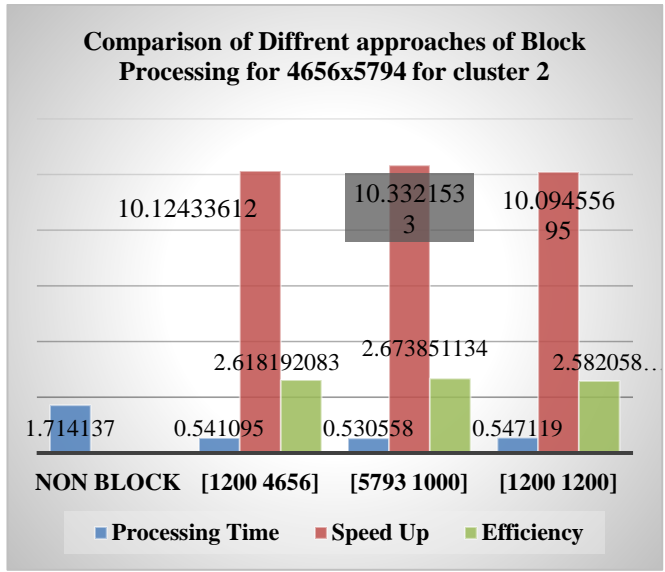

Fig 19. Comparison of Different Approaches of Block Processing for Cluster 2

Table 16 to Table 18 illustrates the results of Sequential with parallel considering the number of cores/workers as 2, 4 and 8 for the image of resolution 4656 x 5793 for Row-Shaped, Column-Shaped and Square Block for Cluster 4.

Table 16. Comparison results of Row-Shaped

| Data Size | Serial | Cores | Row-Shaped | Speed Up | Efficiency |
|---|---|---|---|---|---|
| 4656x5793 | 2.767155 | 2 | 0.249265 | 11.1012577 | 5.550628849 |
| 4656x5793 | 2.767155 | 4 | 0.146973 | 18.82764181 | 4.706910453 |
| 4656x5793 | 2.767155 | 8 | 0.144857 | 19.10266677 | 2.387833346 |

Table 17. Comparison results of Column-Shaped

| Data Size | Serial | Cores | Column-Shaped | Speed Up | Efficiency |
|---|---|---|---|---|---|
| 4656x5793 | 2.767155 | 2 | 0.244717 | 11.3075716 | 5.653785802 |
| 4656x5793 | 2.767155 | 4 | 0.140939 | 19.63370678 | 4.908426695 |
| 4656x5793 | 2.767155 | 8 | 0.144902 | 19.09673434 | 2.387091793 |

Table 18. Comparison results of Square Block

| Data Size | Serial | Cores | Square Block | Speed Up | Efficiency |
|---|---|---|---|---|---|
| 4656x5793 | 2.767155 | 2 | 0.256567 | 10.7853114 | 5.39265572 |
| 4656x5793 | 2.767155 | 4 | 0.14723 | 18.79477688 | 4.69869422 |
| 4656x5793 | 2.767155 | 8 | 0.143322 | 19.30725918 | 2.413407397 |

Table 19 illustrates the comparison results of different approaches of Block processing for cluster 4 followed by its graphical representation.

Table 19. Comparison of Different Approaches of Block processing for cluster 4

| Block Size | Non Block | [1200 4656] Row-Shaped | [5793 1000] Column-Shaped | [1200 1200] Square Block |
|---|---|---|---|---|
| Processing Time | 2.767155 | 0.541095 | 0.530558 | 0.547119 |
| Speed Up | | 49.03156628 | 50.03801273 | 48.88734 |
| Efficiency | | 12.64537265 | 12.94930429 | 12.50475 |

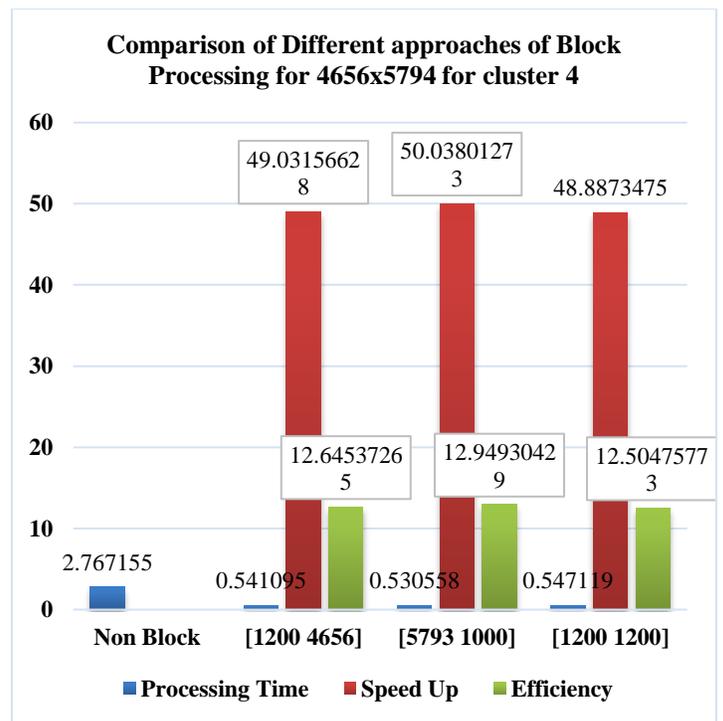

Fig 20. Comparison of Different Approaches of Block Processing for Cluster 4

5. CONCLUSION AND FUTURE SCOPE

This paper mainly focuses on non-hierarchical based K-Means clustering algorithm using Task-level parallelism. This paper mainly focuses on the different approaches of parallel block processing. Row-shaped, column-shaped and square-shaped are the three approaches of parallel block processing. It is experimentally evident that the performance of execution depends mainly on the selection of block size for processing large images and is estimated that column-shaped block size selection is considered to the best among the approaches. Experiment is carried by varying the number of cores as 2, 4, 8 for a different number of clusters as 2, 4 and its performance is evaluated by plotting the graph.

Further the above described approach can be applied for the classification of multispectral images in co-processor and GPU.